\documentclass[ALICE,manyauthors]{cernphprep}
\usepackage[comma,square,numbers,sort&compress]{natbib}

\usepackage{lineno}
\usepackage{xspace}
\usepackage{hyperref}

\usepackage[dvipsnames]{xcolor}
\usepackage{booktabs}
\usepackage{placeins}
\usepackage{enumitem}
\usepackage{amssymb}
\usepackage[T1]{fontenc}
\usepackage[normalem]{ulem}
\modulolinenumbers[5]

\begin{document}
%

\newcommand{\pp}           {pp\xspace}
\newcommand{\ppbar}        {\mbox{$\mathrm {p\overline{p}}$}\xspace}
\newcommand{\XeXe}         {\mbox{Xe--Xe}\xspace}
\newcommand{\PbPb}         {\mbox{Pb--Pb}\xspace}
\newcommand{\pA}           {\mbox{pA}\xspace}
\newcommand{\pPb}          {\mbox{p--Pb}\xspace}
\newcommand{\AuAu}         {\mbox{Au--Au}\xspace}
\newcommand{\dAu}          {\mbox{d--Au}\xspace}

\newcommand{\s}            {\ensuremath{\sqrt{s}}\xspace}
\newcommand{\snn}          {\ensuremath{\sqrt{s_{\mathrm{NN}}}}\xspace}
\newcommand{\pt}           {\ensuremath{p_{\rm T}}\xspace}
\newcommand{\meanpt}       {$\langle p_{\mathrm{T}}\rangle$\xspace}
\newcommand{\ycms}         {\ensuremath{y_{\rm CMS}}\xspace}
\newcommand{\ylab}         {\ensuremath{y_{\rm lab}}\xspace}
\newcommand{\etarange}[1]  {\mbox{$\left | \eta \right |~<~#1$}}
\newcommand{\yrange}[1]    {\mbox{$\left | y \right |~<~#1$}}
\newcommand{\dndy}         {\ensuremath{\mathrm{d}N_\mathrm{ch}/\mathrm{d}y}\xspace}
\newcommand{\dndeta}       {\ensuremath{\mathrm{d}N_\mathrm{ch}/\mathrm{d}\eta}\xspace}
\newcommand{\avdndeta}     {\ensuremath{\langle\dndeta\rangle}\xspace}
\newcommand{\dNdy}         {\ensuremath{\mathrm{d}N_\mathrm{ch}/\mathrm{d}y}\xspace}
\newcommand{\Npart}        {\ensuremath{N_\mathrm{part}}\xspace}
\newcommand{\Ncoll}        {\ensuremath{N_\mathrm{coll}}\xspace}
\newcommand{\dEdx}         {\ensuremath{\textrm{d}E/\textrm{d}x}\xspace}
\newcommand{\RpPb}         {\ensuremath{R_{\rm pPb}}\xspace}

\newcommand{\nineH}        {$\sqrt{s}~=~0.9$~Te\kern-.1emV\xspace}
\newcommand{\seven}        {$\sqrt{s}~=~7$~Te\kern-.1emV\xspace}
\newcommand{\twoH}         {$\sqrt{s}~=~0.2$~Te\kern-.1emV\xspace}
\newcommand{\twosevensix}  {$\sqrt{s}~=~2.76$~Te\kern-.1emV\xspace}
\newcommand{\five}         {$\sqrt{s}~=~5.02$~Te\kern-.1emV\xspace}
\newcommand{\twosevensixnn}{$\sqrt{s_{\mathrm{NN}}}~=~2.76$~Te\kern-.1emV\xspace}
\newcommand{\fivenn}       {$\sqrt{s_{\mathrm{NN}}}~=~5.02$~Te\kern-.1emV\xspace}
\newcommand{\LT}           {L{\'e}vy-Tsallis\xspace}
\newcommand{\GeVc}         {Ge\kern-.1emV/$c$\xspace}
\newcommand{\MeVc}         {Me\kern-.1emV/$c$\xspace}
\newcommand{\TeV}          {Te\kern-.1emV\xspace}
\newcommand{\GeV}          {Ge\kern-.1emV\xspace}
\newcommand{\MeV}          {Me\kern-.1emV\xspace}
\newcommand{\GeVmass}      {Ge\kern-.2emV/$c^2$\xspace}
\newcommand{\MeVmass}      {Me\kern-.2emV/$c^2$\xspace}
\newcommand{\lumi}         {\ensuremath{\mathcal{L}}\xspace}

\newcommand{\ITS}          {\rm{ITS}\xspace}
\newcommand{\TOF}          {\rm{TOF}\xspace}
\newcommand{\ZDC}          {\rm{ZDC}\xspace}
\newcommand{\ZDCs}         {\rm{ZDCs}\xspace}
\newcommand{\ZNA}          {\rm{ZNA}\xspace}
\newcommand{\ZNC}          {\rm{ZNC}\xspace}
\newcommand{\SPD}          {\rm{SPD}\xspace}
\newcommand{\SDD}          {\rm{SDD}\xspace}
\newcommand{\SSD}          {\rm{SSD}\xspace}
\newcommand{\TPC}          {\rm{TPC}\xspace}
\newcommand{\TRD}          {\rm{TRD}\xspace}
\newcommand{\VZERO}        {\rm{V0}\xspace}
\newcommand{\VZEROA}       {\rm{V0A}\xspace}
\newcommand{\VZEROC}       {\rm{V0C}\xspace}
\newcommand{\AD}           {\rm{AD}\xspace}
\newcommand{\ADA}          {\rm{ADA}\xspace}
\newcommand{\ADC}          {\rm{ADC}\xspace}
\newcommand{\Vdecay} 	   {\ensuremath{V^{0}}\xspace}

\newcommand{\ee}           {\ensuremath{e^{+}e^{-}}} 
\newcommand{\pip}          {\ensuremath{\pi^{+}}\xspace}
\newcommand{\pim}          {\ensuremath{\pi^{-}}\xspace}
\newcommand{\kap}          {\ensuremath{\rm{K}^{+}}\xspace}
\newcommand{\kam}          {\ensuremath{\rm{K}^{-}}\xspace}
\newcommand{\pbar}         {\ensuremath{\rm\overline{p}}\xspace}
\newcommand{\kzero}        {\ensuremath{{\rm K}^{0}_{\rm{S}}}\xspace}
\newcommand{\lmb}          {\ensuremath{\Lambda}\xspace}
\newcommand{\almb}         {\ensuremath{\overline{\Lambda}}\xspace}
\newcommand{\Om}           {\ensuremath{\Omega^-}\xspace}
\newcommand{\Mo}           {\ensuremath{\overline{\Omega}^+}\xspace}
\newcommand{\X}            {\ensuremath{\Xi^-}\xspace}
\newcommand{\Ix}           {\ensuremath{\overline{\Xi}^+}\xspace}
\newcommand{\Xis}          {\ensuremath{\Xi^{\pm}}\xspace}
\newcommand{\Oms}          {\ensuremath{\Omega^{\pm}}\xspace}
\newcommand{\degree}       {\ensuremath{^{\rm o}}\xspace}

\newcommand{\pttwo}        {\ensuremath{p^{2}_{\rm T}}\xspace}
\newcommand{\mant}         {\ensuremath{|t|}\xspace}
\newcommand{\GeVtwo}       {Ge\kern-.2emV$^2$/$c^2$\xspace}
\newcommand{\GeVtwoInv}    {$c^2$/Ge\kern-.2emV$^2$\xspace}
\newcommand{\GeVBra}       {[Ge\kern-.2emV]}
\newcommand{\GeVptBra}     {[Ge\kern-.2emV/$c$]}
\newcommand{\GeVmassBra}   {[Ge\kern-.2emV/$c^2$]}
\newcommand{\GeVtwoBra}    {[Ge\kern-.2emV$^2$/$c^2$]}
\newcommand{\slfrac}[2]    {\ensuremath{\left.#1\right/#2}\xspace}
\newcommand{\Ppsi}         {\ensuremath{\rm{\psi'}}\xspace}
\newcommand{\Jpsi}         {\ensuremath{\rm{J/\psi}}\xspace}
\newcommand{\Noon}         {\textbf{n$\mathbf{_O^O}$n}}
\newcommand{\mumu}         {\ensuremath{\mu^{+} \mu^{-}}\xspace}
\newcommand{\elel}         {\ensuremath{e^{+} e^{-}}\xspace}
\newcommand{\pbp}          {\ensuremath{\rm{p \overline{p}}}\xspace}
\newcommand{\eepp}         {\ensuremath{e^{+} e^{-} \pi^{+} \pi^{-}}\xspace}
\newcommand{\mmpp}         {\ensuremath{\mu^{+} \mu^{-} + \pi^{+} \pi^{-}(\pi^{0} \pi^{0})}\xspace}
\newcommand{\llpp}         {\ensuremath{l^{+} l^{-} \pi^{+} \pi^{-}}\xspace}
\newcommand{\lplp}         {\ensuremath{l^{+} l^{-}}\xspace}
\newcommand{\gag}          {\ensuremath{\gamma\gamma}\xspace}

\newcommand{\ptdimuon}     {\ensuremath{p^{\mu\mu}_{\rm T}}\xspace}
\newcommand{\BR}           {\ensuremath{\rm BR}\xspace}
\newcommand{\axe}          {\ensuremath{\rm{Acc\times\epsilon}}\xspace}
\newcommand{\axeVM}        {\ensuremath{\rm{(Acc\times\epsilon)_{VM}}}\xspace}
\newcommand{\invmbarn}     {\ensuremath{\rm{\mu b^{-1}}}\xspace}
\newcommand{\fI}           {\ensuremath{f_{\rm I}}\xspace}
\newcommand{\fD}           {\ensuremath{f_{\rm D}}\xspace}
\newcommand{\Ra}           {\ensuremath{R_{\rm A}}\xspace}
\newcommand{\VMDa}         {\ensuremath{a}\xspace}

\begin{titlepage}
\PHyear{2021}       
\PHnumber{003}      
\PHdate{5 January}  

\title{First measurement of the ${\mathbf{|\textit t|}}$-dependence of coherent $\mathbf{\rm{J/\psi}}$ photonuclear production}
\ShortTitle{First measurement of the $|t|$-dependence of coherent $\mathbf{\rm{J/\psi}}$ photonuclear production}   

\Collaboration{ALICE Collaboration\thanks{See Appendix~\ref{app:collab} for the list of collaboration members}}
\ShortAuthor{ALICE Collaboration} 

\begin{abstract}
The first measurement of the cross section for coherent \Jpsi photoproduction as a function of \mant, the square of the momentum transferred between the incoming and outgoing target nucleus, is presented.
The data were measured with the ALICE detector in ultra-peripheral \PbPb collisions at a centre-of-mass energy per nucleon pair \fivenn with the \Jpsi produced in the central rapidity region $|y|<0.8$, which corresponds to the small Bjorken-$x$ range  $(0.3-1.4)\times10^{-3}$. 

The measured \mant-dependence is not described by computations based only on the Pb nuclear form factor, while the photonuclear cross section is better reproduced by models including shadowing according to the leading-twist approximation, or gluon-saturation effects from the impact-parameter dependent Balitsky--Kovchegov equation. These new results are therefore a valid tool to constrain the relevant model parameters and to investigate the transverse gluonic structure at very low \mbox{Bjorken-$x$}.

\end{abstract}
\end{titlepage}

\setcounter{page}{2} 


\section{Introduction} 

Photonuclear reactions can be studied in  ultra-peripheral collisions (UPCs) of heavy ions where the two projectiles pass each other with an impact parameter larger than the sum of their radii. In this case, purely hadronic interactions are suppressed and electromagnetically induced processes occur via photons with typically very small virtualities, of the order of tens of MeV$^2$. The intensity of the photon flux is proportional to the square of the electric charge of the nuclei, resulting in large cross sections for the coherent photoproduction of a vector meson in UPCs of Pb ions at the LHC. This process has a clear experimental signature: the decay products of the vector meson are the only particles detected in an otherwise empty detector.

The physics of vector meson photoproduction is described, e.g., in Refs.~\cite{Bertulani:2005ru, Baltz:2007kq,Contreras:2015dqa,Klein:2019qfb}. Two vector meson photoproduction processes, coherent and  incoherent, are relevant for the results presented here. In the former, the photon interacts with all nucleons in a nucleus, while in the latter it interacts with a single nucleon. In both cases a single vector meson is produced. Experimentally, one can distinguish between these two production types through the transverse momentum \pt of the vector meson which is related to the transverse size of the target. While coherent photoproduction is characterised by an average transverse momentum $ \left<\pt\right> \sim$~60~MeV/$c$, incoherent production leads to higher average transverse momenta:~$ \left<\pt\right> \sim$~500~MeV/$c$. Incoherent photoproduction can also be accompanied by the excitation and dissociation of the target nucleon resulting in an  even higher transverse momentum of the produced vector meson~\cite{Guzey:2018tlk}.

Shadowing, the observation that the structure of a nucleon inside nuclear matter is different from that of a free nucleon~\cite{Armesto:2006ph}, is not yet completely understood and several processes may have a role in different kinematic regions. In this context, coherent heavy vector meson photoproduction is of particular interest, because  it is especially sensitive to the gluon distribution in the target, and thus to gluon shadowing effects at low Bjorken-$x$~\cite{Ryskin:1992ui,Rebyakova:2011vf}. One of the effects expected to contribute to shadowing in this kinematic region is saturation, a dynamic equilibrium between gluon radiation and recombination~\cite{Albacete:2014fwa}. The momentum scale of the interaction ($Q^{2}$) is related to the mass $m_V$ of the vector meson as $Q^{2}~\sim~m^{2}_{V}/4$, corresponding to the perturbative regime of quantum chromodynamics (QCD) in the case of  charmonium states. The rapidity of the coherently produced ${\rm c} \bar{\rm c}$ states is related to the Bjorken-$x$ of the gluonic exchange as $x~=~\left(m_V/\sqrt{s_{\rm NN}}\right)\exp\left(\pm~y\right)$, where the two signs indicate that either of the incoming ions can be the source of the photon. Thus, the charmonium photoproduction cross section at midrapidity in \PbPb UPCs at the LHC Run~2 centre-of-mass energy per nucleon pair of \fivenn is sensitive to $x\in (0.3,1.4)\times10^{-3}$ at ALICE. It thereby provides information on  the gluon distribution in nuclei in a kinematic region where shadowing could be present and saturation effects may be important~\cite{Guzey:2016qwo,Bendova:2020hbb}.

Charmonium photoproduction in ultra-peripheral \PbPb collisions was previously studied by the ALICE Collaboration at \twosevensixnn~\cite{Abelev:2012ba, Abbas:2013oua,Adam:2015sia}. The coherent \Jpsi photoproduction cross section was measured both at midrapidity $|y|<0.9$ and at forward rapidity $-3.6 < y < -2.6$.  Recently, a measurement of the rapidity dependence of coherent \Jpsi  photoproduction at forward rapidity at the higher energy of \fivenn was also published by the ALICE Collaboration~\cite{Acharya:2019vlb}. In addition, the CMS Collaboration studied the coherent \Jpsi\ photoproduction accompanied by neutron emission at semi-forward rapidity $1.8 < |y| < 2.3$ at \twosevensixnn~\cite{Khachatryan:2016qhq}. These measurements allow for a deeper insight into the rapidity dependence of gluon shadowing, but do not give information on the behaviour of gluons in the impact-parameter plane.
The square of the momentum transferred to the target nucleus, \mant, is related through a two-dimensional Fourier transform to the gluon distribution in the plane transverse to the interaction~\cite{Bartels:2003yj}; thus the study of the \mant-dependence of coherent \Jpsi\ photoproduction provides information about the spatial distribution of gluons as a function of the impact parameter.
Thus far, the only measurements in this direction were performed recently by the STAR Collaboration for the case of the $\rho^{0}$ vector meson~\cite{Adamczyk:2017vfu} and for the yield of \Jpsi in semi-central Au--Au collisions~\cite{STAR:2019yox}.

In this Letter, the first measurement of the \mant-dependence of the coherent \Jpsi\ photoproduction cross section at midrapidity in \PbPb UPCs at \fivenn is presented. The \Jpsi vector mesons were reconstructed in the rapidity range $|y|<0.8$ through their decay into \mumu, taking advantage of the better mass and momentum resolution of this channel with respect to the $e^+e^-$ channel. The data sample, recorded in 2018,  is approximately 10 times larger than that used in  previous ALICE measurements at midrapidity at the lower energy of \twosevensixnn \cite{Adam:2015sia}.  Cross sections are reported for six \mant intervals and  compared with  theoretical predictions.

\section{Detector description}

The ALICE detector and its performance are described in Refs.~\cite{Aamodt:2008zz, Abelev:2014ffa}. Three central barrel detectors, the Inner Tracking System (\ITS), the Time Projection Chamber (\TPC), and the Time-of-Flight (\TOF), in addition to two forward detectors, \VZERO and the ALICE Diffractive (\AD) arrays, are used in this analysis. The central barrel detectors are surrounded by a large solenoid magnet producing a magnetic field of $B = 0.5$~T. The \VZERO, \AD, \ITS, and \TOF detectors are used for triggering, the \ITS and the \TPC for particle tracking, and the \TPC for particle identification.

The \VZERO is a scintillator detector made of two counters, V0A and V0C, installed on both sides of the interaction point. The V0A and V0C cover the pseudorapidity ranges $2.8< \eta <5.1$ and $-3.7< \eta <-1.7$, respectively. Both counters are segmented in four rings in the radial direction, with each ring divided into 8 sections in azimuth.

The \AD consists of two scintillator stations, ADA and ADC, located at 16 and $-19$ m along the beam line with respect to the nominal interaction point and covering the pseudorapidity ranges $4.8< \eta <6.3$ and $-7.0 < \eta <-4.9$, respectively~\cite{Akiba:2016ofq,Broz:2020ejr}.

The \ITS is a silicon based detector and is made of six cylindrical layers using three different technologies. The Silicon Pixel Detector (\SPD) forms the two innermost layers of the ITS and covers $|\eta|<2$ and $|\eta|<1.4$, respectively. Apart from tracking, the \SPD is also used for triggering purposes and to reconstruct the primary vertex.

The \ITS is cylindrically surrounded by the \TPC, whose main purpose is to track particles and provide charged-particle momentum measurements with good two-track separation and particle identification. The TPC coverage in pseudorapidity is $|\eta|<0.9$ for tracks with full radial length.   The TPC has full coverage in azimuth. It offers good momentum resolution in a large range of the track transverse momentum spanning from 0.1~GeV/$c$ to 100~GeV/$c$.

The \TOF is a large cylindrical gaseous detector based on multi-gap resistive-plate chambers. It covers the pseudorapidity region $|\eta|<0.8$. The \TOF readout channels are arranged into 18 azimuthal sectors which can provide topological trigger decisions.

\section{Data analysis}

\subsection{Event selection} \label{sec_event_selection}

The online event selection was based on a dedicated UPC trigger which selected back-to-back tracks in an otherwise empty detector. This selection required ($i$) that nothing above the trigger threshold was detected in the \VZERO and \AD detectors, ($ii$) a topological trigger requiring less than eight \SPD chips with trigger signal, forming at least two pairs; each pair was required to have an \SPD chip fired in each of the two layers and to be in compatible azimuthal sectors, with an opening angle in azimuth between the two pairs larger than 144\degree, ($iii$) a topological trigger in the \TOF requiring more than one and less than seven \TOF sectors to register a signal; at least two of these sectors should have an opening angle in azimuth larger than 150\degree.

The integrated luminosity of the analysed sample is 233~$\mu\text{b}^{-1}$. The determination of the luminosity is obtained from the counts of a reference trigger based on multiplicity selection in the \VZERO detector, with the corresponding cross section estimated from a van der Meer scan; this procedure has an uncertainty of 2.2$\%$~\cite{ALICE-PUBLIC-2021-001}. The determination of the live-time of the UPC trigger has an additional uncertainty of 1.5$\%$. The total relative systematic uncertainty of the integrated luminosity is thus 2.7$\%$.

Additional offline \VZERO and \AD veto decisions were applied in the analysis. The offline veto algorithm improved the signal to background ratio, because it utilised a larger timing window to integrate the signal than its online counterpart. Some good events were lost due to this selection. The loss was taken into account with the correction on veto trigger inefficiency discussed in Sec.~\ref{sec_axe}. The systematic uncertainty from the \VZERO and \AD vetoes was estimated as the relative change in the measured \Jpsi cross section before and after imposing them and correcting for the losses; it amounts to 3\%.

Each event had a reconstructed primary vertex within 15~cm from the nominal interaction point along the beam direction, $z$, and had exactly two tracks. These tracks were reconstructed using combined tracking in the \ITS and \TPC. Tracks were requested to have at least 70 (out of 159) \TPC space points and to have a hit in each of the two layers of the \SPD. Each track had to have a distance of closest approach to the event interaction vertex of less than 2 cm in the $z$-axis direction. Also, each track was required to have $|\eta|<0.9$. The relative systematic uncertainty from tracking, which takes into account the track quality selection and the track propagation from the \TPC to the \ITS, was estimated from a comparison of data and Monte Carlo simulation. The combined uncertainty to reconstruct both tracks is 2.8\%.

The particle identification (PID) was provided by the specific ionisation losses in the \TPC, which offer a large separation power between muons and electrons from the leptonic decays of the \Jpsi in the momentum range $(1.0,2.0)$~\GeVc, relevant for this analysis. The effect of a possible misidentification was found to be negligible.

An offline \SPD decision was also applied in the analysis. The offline topological \SPD algorithm ensured that the selected tracks crossed the \SPD chips used in the trigger decision. The relative systematic uncertainty from the \SPD and \TOF trigger amounts to 1.3$\%$, which was estimated using a data-driven method by changing the requirements on the probe tracks.

The selected events were required to have tracks with opposite electric charge, the rapidity of the dimuon candidate was restricted to $|y|<0.8$ and its \pt had to be less than 0.11~\GeVc, in order to obtain a sample dominated by coherent interactions with just a small contamination from incoherent processes. The measurement was initially carried out in \pttwo intervals, because for collider kinematics \mant$\approx~$\pttwo. The corrections needed to obtain the \mant-dependence are discussed in  Sec.~\ref{sec:pt2tot}.

\subsection{Signal extraction}

As a first step in extracting the coherent \Jpsi\ signal, a fit to the opposite sign dimuon invariant mass distribution was performed. The model used to fit the data consists of three templates: one Crystal Ball function~\cite{Oreglia:1980cs} (CB) to describe the \Jpsi resonance, a second CB function to describe the \Ppsi resonance, and an exponential function to describe the continuum production of muon pairs, $\gamma\gamma\to\mu^+\mu^-$. 

The parameters of the exponential function  were left free. The integral of this exponential in the mass range $(3.0,3.2)$ \GeVmass was used to determine the number of  events from the continuum production in this interval.

The CB parameters describing the tails of the measured distribution in data, commonly known as $\alpha$ and $n$, were fixed to the values obtained while fitting the dimuon invariant mass distribution in an associated Monte Carlo simulation, which is described in Sec.~\ref{sec_axe}. These settings were employed for both  CB functions. 

\begin{figure}[!th]
    \begin{center}
        \includegraphics[width=0.46\linewidth]{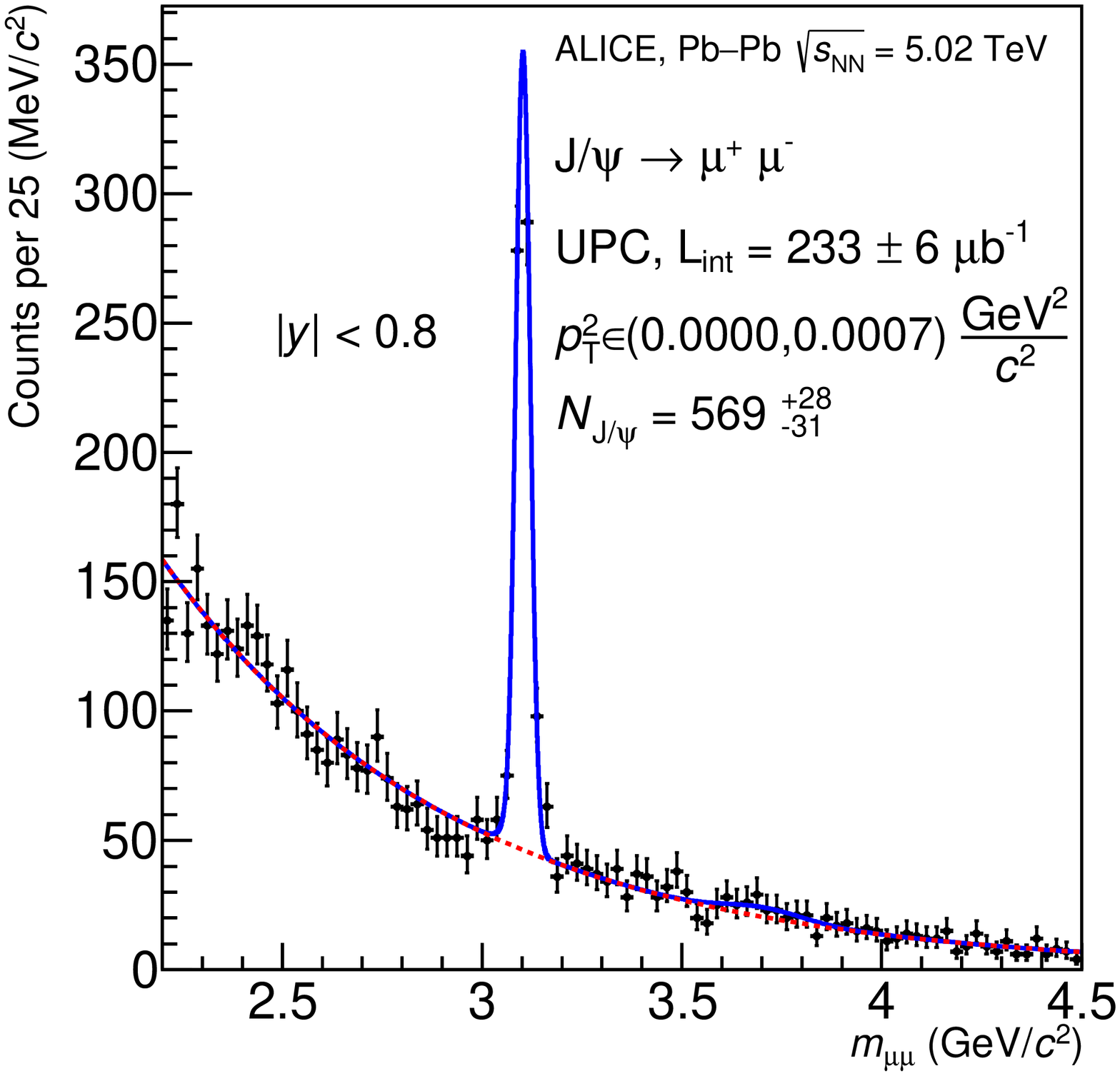}
        \includegraphics[width=0.46\linewidth]{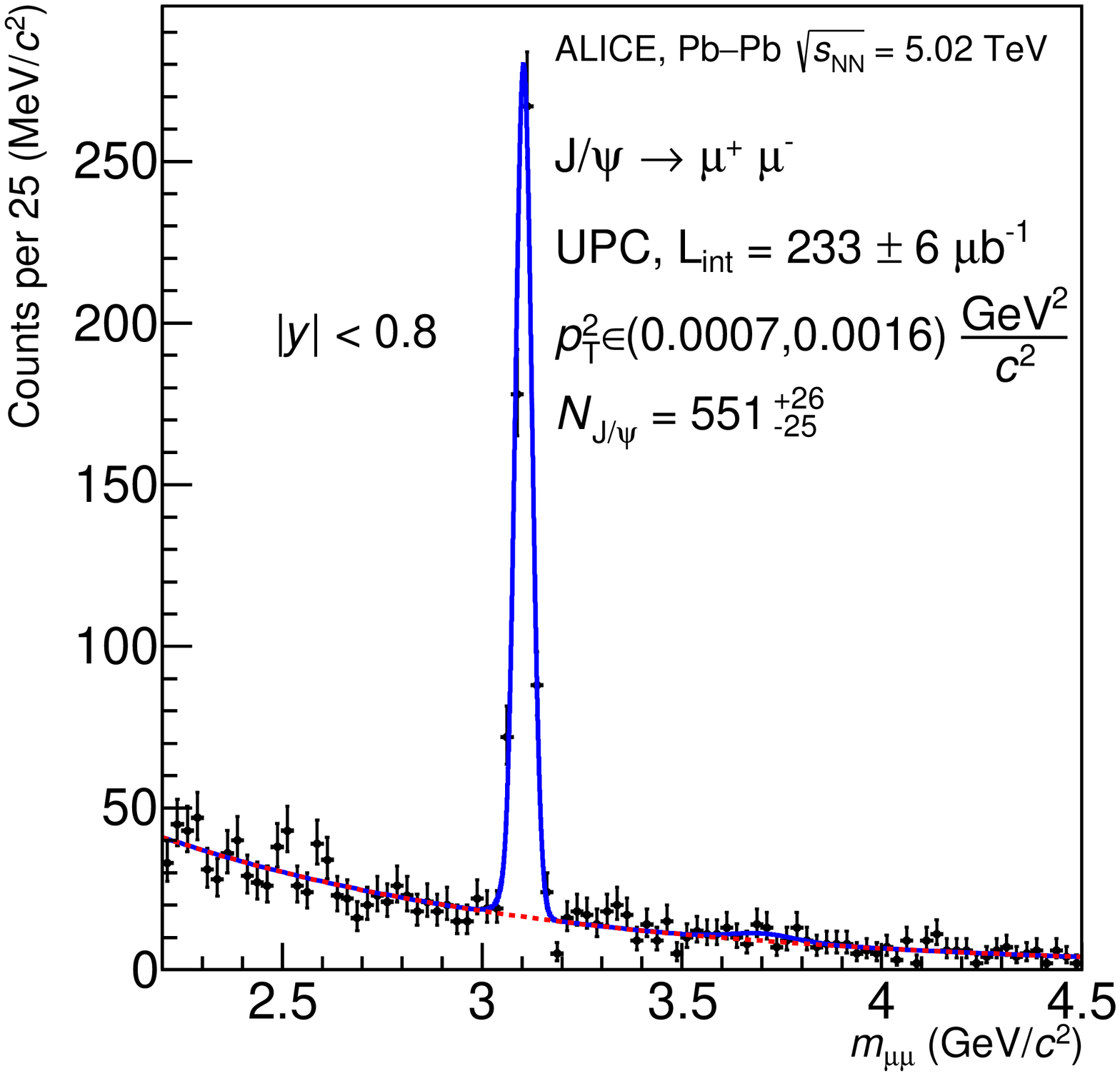}\\
        \includegraphics[width=0.46\linewidth]{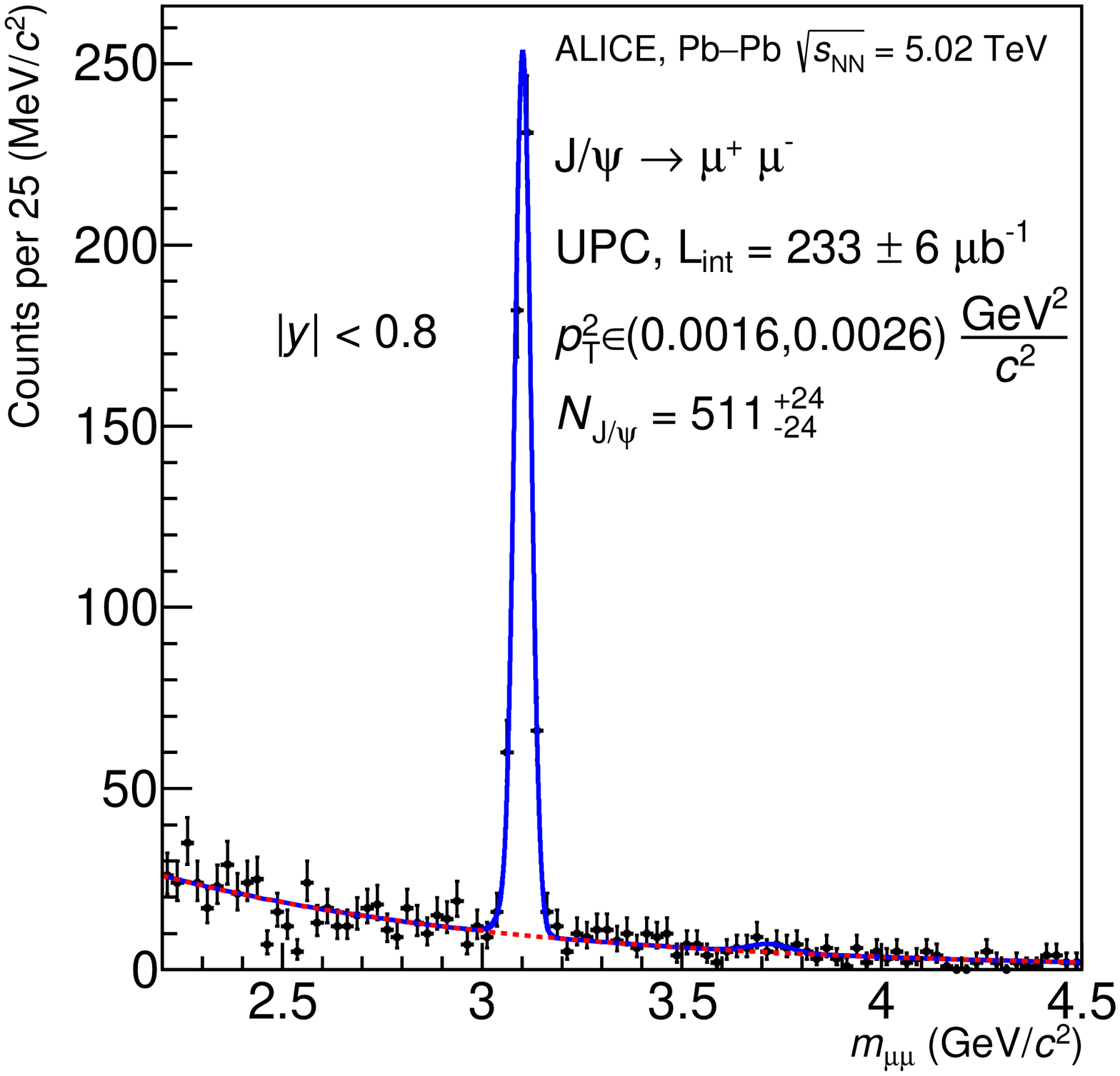}
        \includegraphics[width=0.46\linewidth]{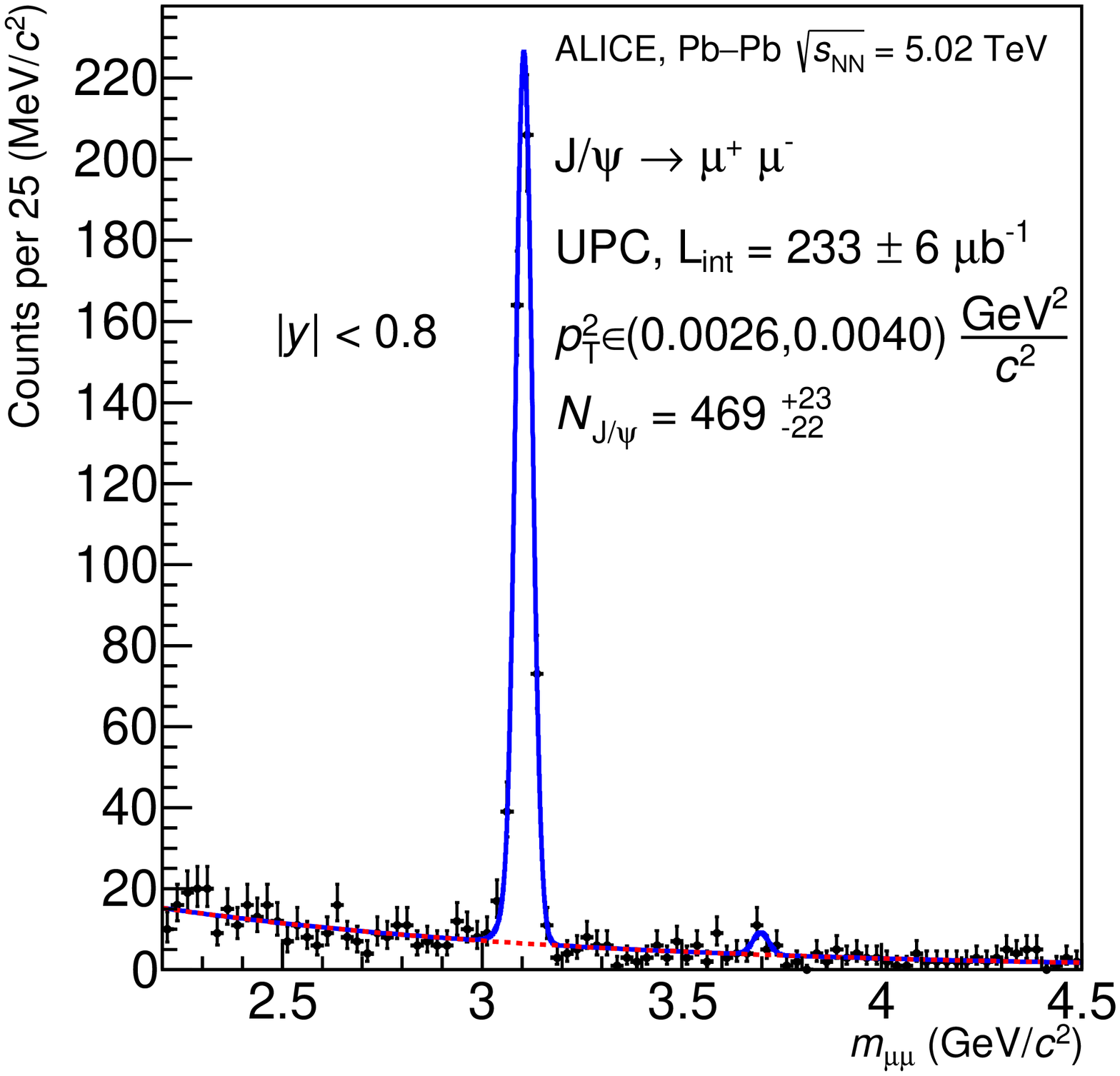}\\
        \includegraphics[width=0.46\linewidth]{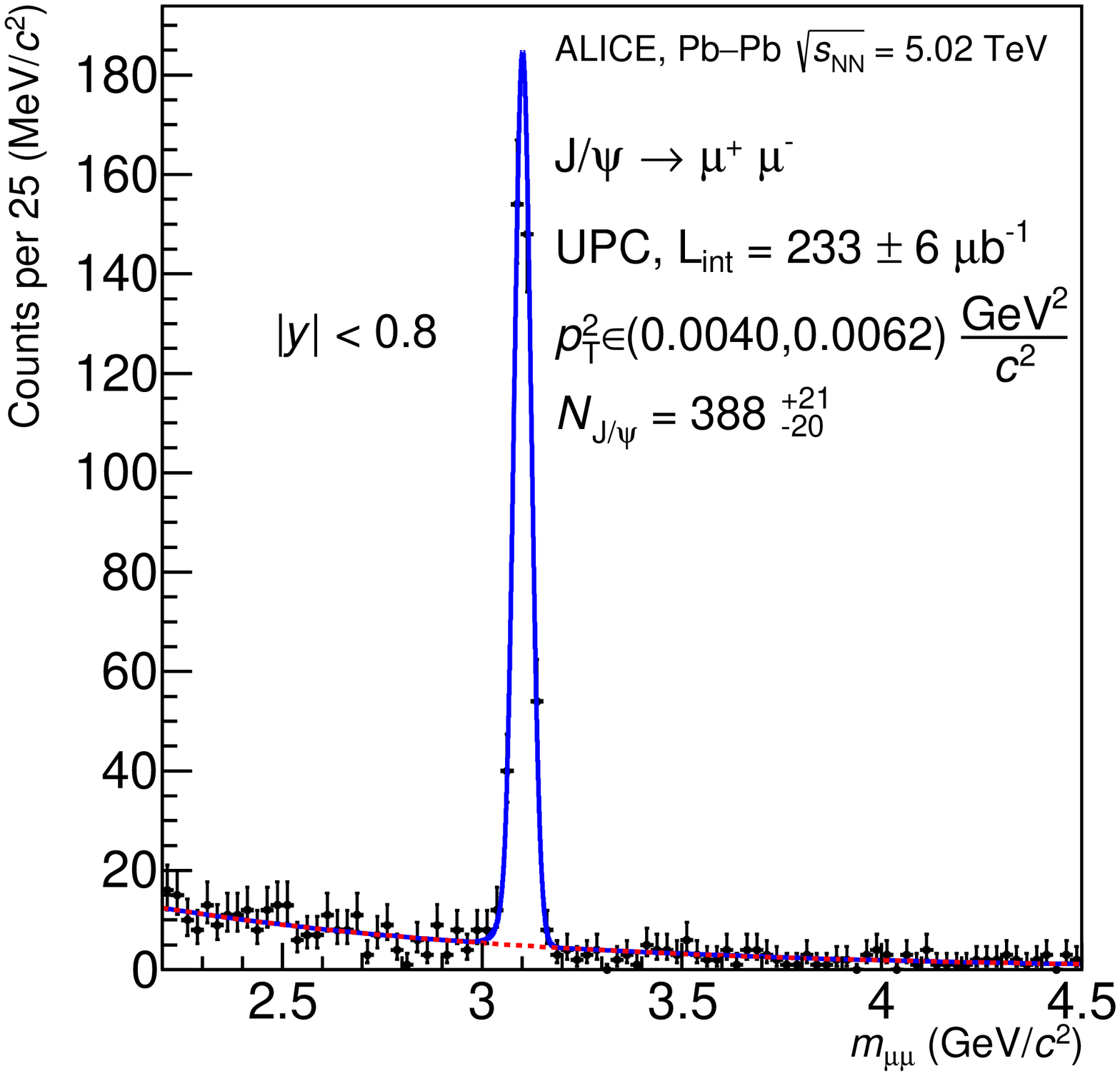}
        \includegraphics[width=0.46\linewidth]{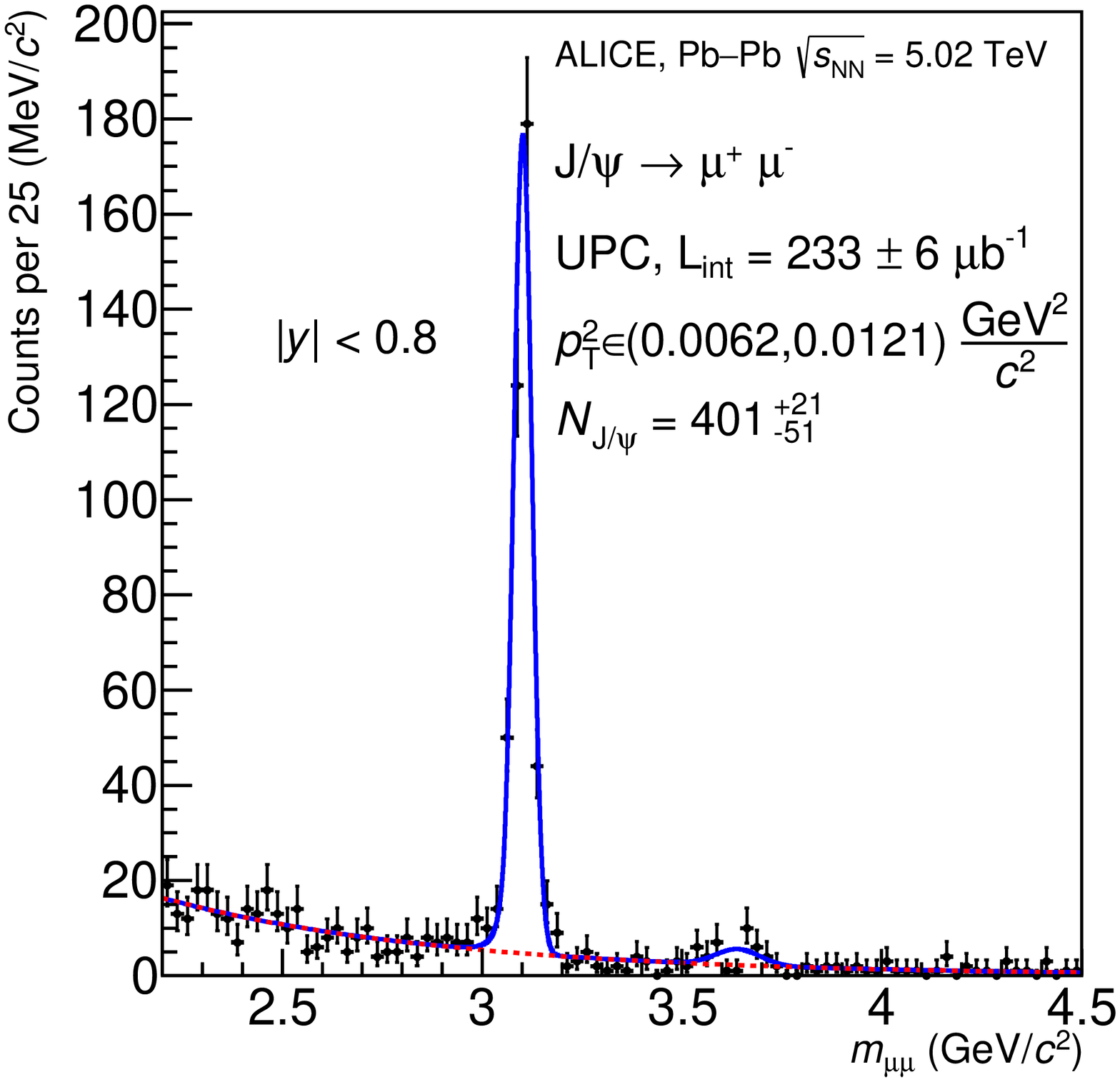}
    \end{center}
    \caption{Invariant-mass distributions for different \pttwo intervals with the global fit described in the text shown with the blue line. The exponential part of the fit model, representing the $\gamma\gamma\to\mu^+\mu^-$ background, is shown in red.}
    \label{fig_massNbins}
\end{figure}

The number of \Jpsi candidates in each \pttwo interval was obtained from an extended maximum likelihood fit to the unbinned invariant mass distribution of all $\mu^+\mu^-$ pairs  which survived the selection criteria described in Sec.~\ref{sec_event_selection}. Results of the fits for the six \pttwo intervals are shown in Fig.~\ref{fig_massNbins}. In all cases a very clear \Jpsi resonance is seen over a fairly small background. Note that the effect on the kinematics from a potential dimuon decay including bremsstrahlung is negligible.

The relative systematic uncertainty from the signal extraction was calculated by repeating the fit over different invariant mass ranges, and modifying the CB $\alpha$ and $n$ parameters accordingly. These uncertainties vary in the interval (0.7,2.2)\%.
 
\subsection{Corrections for irreducible backgrounds}

The selection criteria described above are not sensitive to events which mimic the signature of coherent \Jpsi production, but are coming from feed-down of \Ppsi or incoherent production. The contribution of these events was taken into account with the  \fD and \fI factors, respectively, entering Eq.~(\ref{eq_yieldCorrection}), 

\begin{equation} \label{eq_yieldCorrection}
 N^{\text{coh}}_{\Jpsi} = \frac{N^{\text{fit}}}{1 + \fI + \fD}\times\frac{1}{(\axe)^{\text{coh}}_{\Jpsi}},
\end{equation}
where $N^{\text{fit}}$, the yield of \Jpsi candidates, is the integral of the CB describing the \Jpsi signal in the fit of the dimuon invariant mass spectrum, and $(\axe)^{\text{coh}}_{\Jpsi}$ is the acceptance and efficiency correction factor described in Sec.~\ref{sec_axe}. 

Feed-down refers to the decay of a \Ppsi to a \Jpsi plus anything else, where these additional particles were not detected for some reason. The correction for these  events, \fD, was estimated with Monte Carlo simulations describing the apparatus (\axe) factor for the following channels: \Jpsi$\rightarrow$\mumu, \Ppsi$\rightarrow$\mumu, and \Ppsi$\rightarrow \Jpsi + X$; and the measured ratio of \Ppsi to \Jpsi production cross sections. The details of the method are described in Ref.~\cite{Acharya:2019vlb}. The results for each \pttwo interval are summarised in Table~\ref{tab_yieldcorrection}. Relative systematic uncertainties, estimated by using different cross section ratios, are \pttwo-correlated. Their relative effect on the final cross section can be found in Table~\ref{tab_syserrcorrelation}; it is well below 1\%.

Most of the incoherent production of \Jpsi off nucleons was rejected with the restriction of the phase space in \pt, as mentioned in Sec.~\ref{sec_event_selection}. However, around 5\%  of all incoherent events remained in the region where the measurement was performed. To estimate the \fI factor to correct for the remaining incoherent events, a fit to the measured \Jpsi \pt distribution of data in the invariant mass range $(3.0,3.2)$~\GeVmass was used. The model  fitted to the data consists of six templates: coherent~\Jpsi photoproduction, incoherent~\Jpsi photoproduction, incoherent~\Jpsi photoproduction with nucleon dissociation, coherent~\Ppsi photoproduction, incoherent~\Ppsi photoproduction, and continuum production from $\gamma\gamma\to\mu^+\mu^-$. The templates of all, but dissociative \Jpsi and continuum, were taken from Monte Carlo simulations. In the fit, the fractions of both \Ppsi photoproduction processes were fixed to values calculated as described above. These included the modifications that the \pt restriction was released and that there was a selection on the invariant mass to be in the range $(3.6,3.8)$~\GeVmass. Other fractions were left free in the fit. The normalisation of the continuum was restricted from the invariant mass fit to be the sum of background events in the mass range of the \Jpsi. The shape of the continuum was taken from the dimuon \pt distribution selecting the invariant mass range between the \Jpsi and the \Ppsi, while the shape for the nucleon dissociation process was based on the H1 parameterisation~\cite{Alexa:2013xxa}. The global template was fitted to data using an extended maximum likelihood unbinned fit. The results for each \pttwo interval are reported in Table~\ref{tab_yieldcorrection}. The systematic uncertainties, estimated from a combination of the fit uncertainty and a modification of the coherent template used in the fitting model are \pt-correlated. Their relative effect on the final cross section can be found in Table~\ref{tab_syserrcorrelation}.

\subsection{Acceptance, efficiency and pile-up corrections
\label{sec_axe}}

The STARlight 2.2.0 MC generator~\cite{Klein:2016yzr} was used to generate samples of coherent and incoherent events for the production of \Jpsi$\rightarrow$\mumu and \Ppsi$\rightarrow$\mmpp. GEANT 3.21~\cite{Brun:1082634} was used to reproduce the response of the detector.  The simulated data were reconstructed with the same software as the real ones, accounting for actual data-taking conditions. Values of the acceptance and efficiency, $(\axe)^{\text{coh}}_{\Jpsi}$, are shown in Table~\ref{tab_yieldcorrection} for the different \pttwo intervals used in this analysis. 

\AD and \VZERO were used to veto activity at forward rapidity. These detectors were sensitive to signals coming from independent interactions (pile-up), which  resulted in the rejection of potentially interesting events. The correction factor for this effect was obtained using a control sample of events collected with an unbiased trigger. These were then used to compute the probability of having a veto from \AD or \VZERO in otherwise empty events. The total veto trigger efficiency $\epsilon^{\text{VETO}}$ used in Eq.~(\ref{eq_expcohCS}) was determined to be 0.94. The corresponding systematic uncertainty is included in the \AD and \VZERO value of 3\%  mentioned in Sec.~\ref{sec_event_selection}.

Electromagnetic dissociation (EMD) is another process which may cause the rejection of a good event due to the veto from the forward detectors. EMD can occur when photons excite one or both interacting nuclei. Upon de-excitation, neutrons and sometimes other charged particles are emitted at forward rapidities~\cite{Pshenichnov:1999hw} and can trigger a \VZERO or \AD veto. Such loss of events was quantified from data gathered with a specialized EMD trigger; the efficiency correction factor to take into account these losses amounts to $\epsilon^{\text{EMD}}~=~0.92$ with a relative systematic uncertainty of 2\% given by the statistical uncertainty from the control sample.

\begin{table*}[t]
\caption{Incoherent correction \fI, feed-down correction \fD and the $(\axe)^{\text{coh}}_{\Jpsi}$ correction factor for each \pttwo interval. See  Eq.~(\ref{eq_yieldCorrection}).}
\centering
\begin{tabular}{lccc}
\toprule
    \pttwo interval (\GeVtwo) & \fI & \fD & $(\axe)^{\text{coh}}_{\Jpsi}$\\
    \midrule
    $\left(0,0.00072\right)$ & 0.0045 & 0.0039 & 0.0348\\
    $\left(0.00072,0.0016\right)$ & 0.0047 & 0.0046 & 0.0352\\
    $\left(0.0016,0.0026\right)$ & 0.0047 & 0.0058 & 0.0358\\
    $\left(0.0026,0.004\right)$ & 0.0072 & 0.0072 & 0.0365\\
    $\left(0.004,0.0062\right)$ & 0.0120 & 0.011 & 0.0379\\
    $\left(0.0062,0.0121\right)$ & 0.0300 &  0.028 & 0.0412\\
    \bottomrule
\end{tabular}
\label{tab_yieldcorrection}
\end{table*}

\subsection{Unfolding of the $\mathbf{\textit p^{2}_{\rm T}}$ distribution}

Cross sections were measured in different \pttwo intervals. In order to account for the migration of about 45\% of the events across \pttwo intervals due to the finite resolution of the detector, an unfolding procedure was used. The effect of migrations are much more important than the small difference between the data and MC \pttwo spectra, so no re-weighting has been performed previous to unfolding.

Amongst many available methods, unfolding based on Bayes' theorem~\cite{DAgostini:1994fjx} was chosen to perform the unfolding, while the singular-value decomposition (SVD) method~\cite{Hocker:1995kb} served to study potential systematic effects. The implementations of these methods as provided by RooUnfold~\cite{Adye:2011gm} were used in this analysis.

Bayesian unfolding is an iterative method, therefore the result depends on the number of iterations. The size of the data sample is large enough to investigate different numbers of \pttwo ranges. These two parameters, that is the number of iterations and of ranges, were tuned using Monte Carlo simulations by studying the evolution of the statistical uncertainty in each interval as a function of the number of iterations, and by using the relative difference between iteration-adjacent results. It was found that the best combination for this analysis is Bayes' unfolding with three iterations applied to the \pttwo distribution split into six regions. The widths of the \pttwo intervals were chosen to have similar statistical uncertainties in each region.

The Monte Carlo sample used for unfolding contained 600 000 events. An 80\% fraction of them was used to train the response matrix which is used to unfold the true distribution from the measured distribution. This matrix was tested on the remaining 20\% of the events. The unfolding matrix was able to correct the smeared distribution with high precision. Comparison with results using the SVD method revealed a \pt-correlated  relative systematic uncertainty with values in the interval (0.6,2.3)\%.

\subsection{Cross section for coherent $\mathbf{\rm{J}/\psi}$ photoproduction in UPCs}

The differential cross section for coherent \Jpsi photoproduction in a given \pttwo interval and a given rapidity range $\Delta y$ in Pb--Pb UPCs is 
\begin{equation} \label{eq_expcohCS}
 \frac{{\rm d}^{2}\sigma^{\text{coh}}_{\Jpsi}}{{\rm d}y{\rm d}\pttwo} = \frac{^{\text{unf}}N^{\text{coh}}_{\Jpsi}}{\epsilon^{\text{VETO}}\times\epsilon^{\text{EMD}}\times\BR(\Jpsi\rightarrow\mu^{+}\mu^{-})\times\lumi_{\text{int}}\times\Delta \pttwo\times\Delta y},
\end{equation}
where the correction factors $\epsilon^{\text{VETO}}$ and $\epsilon^{\text{EMD}}$ are introduced in Sec.~\ref{sec_axe}, $\BR(\Jpsi\rightarrow\mu^{+}\mu^{-})$ is the branching ratio ($5.961 \pm 0.033$)$\%$~\cite{Zyla:2020}, $\lumi_{\text{int}}$ is the total integrated luminosity of the data sample, $\Delta \pttwo$ is the size of the interval where the measurement was performed, and finally, $^{\text{unf}}N^{\text{coh}}_{\Jpsi}$ is the number of coherent \Jpsi candidates after unfolding the results given by Eq.~(\ref{eq_yieldCorrection}). The corresponding systematic uncertainties are summarised in the upper part of Table~\ref{tab_syserrcorrelation}.
With the exception of signal extraction, all other systematic uncertainties mentioned up to here are correlated across \pttwo intervals.

\subsection{Corrections for the photonuclear cross section
\label{sec:pt2tot}}

The cross section described by Eq.~(\ref{eq_expcohCS}) is the one measured by ALICE. The main theoretical interest is in the photonuclear process at a fixed energy. To obtain the corresponding cross section, one has to account for several effects. None of these effects is affected by the ALICE detector, they just depend on the kinematics and quantum nature of the process. This means that the uncertainties in going from the UPC to the photonuclear cross sections are of theoretical nature only. 

At midrapidity, the UPC cross section corresponds to the $\gamma$Pb cross section multiplied by twice the photon flux averaged over the impact parameter, $n_{\gamma{\rm Pb}} (y)$,

\begin{equation} 
\label{eq_upc_gPb}
\left. \frac{{\rm d}^{2}\sigma^{\text{coh}}_{\Jpsi}}{{\rm d}y{\rm d}\pttwo}
\right|_{y=0}= 2n_{\gamma{\rm Pb}} (y=0)
\frac{{\rm d}\sigma_{\gamma{\rm Pb}}}{{\rm d}|t|}.
\end{equation}

Since the rapidity dependence of the UPC cross section in the rapidity range studied here is fairly flat, the measurements are taken to represent the value at $y=0$. In UPCs, there are two potential photon sources, so in principle both amplitudes have to be added and their interference needs to be accounted. This was studied for the first time in Ref.~\cite{Klein:1999gv} and later measured for the case of $\rho^0$ coherent photoproduction by the STAR Collaboration~\cite{Abelev:2008ew}. The interference is important only at very small values of \mant (see for example~\cite{Zha:2017jch}). To account for this effect, the STARlight program, which includes the interference of both amplitudes, was used. It was found that this is an 11.6$\%$ effect in the smallest \mant interval, where the effect is concentrated. To estimate the potential uncertainty on this procedure, the interference effects with the nominal strength were compared to those with a 25\% reduction of the strength. The relative change in the photonuclear cross section varied from 0.3 to 1.2\% with the largest uncertainty being assigned to the smallest \mant interval.

The photon flux was computed in the semiclassical formalism following the prescription detailed in Ref.~\cite{Contreras:2016pkc} and cross checked with that of Ref.~\cite{Broz:2019kpl}. The flux amounts to 84.9 with an uncertainty of 2\% coming from variations of the geometry of the Pb ions. 

Although the value of \pttwo  is a good approximation to that of \mant, it is not exact due to the fact that the photon also has a transverse momentum in the laboratory frame. To account for this effect, the cross section was unfolded with a response matrix built from \pttwo- and \mant-distributions. Two sources for the distributions were used: ($i$) the STARlight generator which includes the transverse momenta of the photons, but does not describe so well the shape of the measured \pttwo distribution in data, and ($ii$)  measured \pttwo values coupled to photon momenta randomly generated using the transverse momentum distribution of photons from Refs.~\cite{Vidovic:1992ik,Hencken:1995me}. The average of the corresponding unfolded results was used for the cross section, while half their difference was taken as a systematic uncertainty which varied between 0.1\% and 5.7\%, with this last value corresponding to the largest \mant interval. 

These three uncertainties are reported in the lower part of Table~\ref{tab_syserrcorrelation}. The uncertainty on the value of the photon flux at $y=0$ is correlated across \mant, the uncertainty on the \pttwo$\rightarrow$\mant unfolding is partially correlated and the uncertainty on the variation of the interference term is anti-correlated in the lowest \mant region and correlated in the other \mant regions. They are added in quadrature for the final result shown in Sec.~\ref{sec_res} and Table~\ref{tab_cs} below.

\begin{table}[t]
  \caption{Summary of the identified systematic uncertainties on the coherent \Jpsi photoproduction and photonuclear cross sections. The uncertainties to go from the measured cross section in UPCs to the photonuclear process are listed after the line in the middle of the table and their origin depends on the modeling of the photon flux and interference effects.  The correlation across \pttwo intervals is discussed in the text.}
  \begin{center}
    \begin{tabular}{l c c}
    \toprule
    Source & Uncertainty ($\%$) \\
    \midrule
    Signal extraction & (0.7,2.2)\\
    $f_{\rm D}$  & (0.1,0.5) \\
    $f_{\rm I}$  &(1.1,2.3) \\
    \pttwo migration unfolding  & (0.6,2.3)\\
    Luminosity  & 2.7\\
    V0 and AD veto & 3\\
    EM dissociation  & 2\\
    ITS-TPC tracking  & 2.8\\
    SPD and TOF efficiency  & 1.3\\
    Branching ratio & 0.5\\
    \midrule
    Variations in interference strength & (0.3,1.2)\\
    Value of the photon flux at $y=0$ & 2\\
    \pttwo$\rightarrow$\mant unfolding & (0.1,5.7)\\
    \bottomrule
    \end{tabular}
    \label{tab_syserrcorrelation}
  \end{center}
\end{table}

\section{Results
\label{sec_res}}

The final result for the cross section measured in each \pttwo interval is reported in Table~\ref{tab_cs}. The statistical uncertainty originates from the error obtained in the fit to the dimuon invariant-mass distribution, propagating the uncertainties of the \fI and \fD corrections, see Eq.~(\ref{eq_yieldCorrection}), and the uncertainty related to the unfolding process. The uncorrelated systematic uncertainty from signal extraction and the quadratic sum of correlated systematic uncertainties are shown in Table~\ref{tab_cs}.

\begin{table*}[t]
\caption{Measured coherent \Jpsi photoproduction cross section in UPCs in different $\pttwo$ intervals as well as the photonuclear cross section in \mant-intervals. The first uncertainty is statistical, the second and third systematic, uncorrelated and correlated, respectively. The fourth uncertainty, for the photonuclear cross section case, is the systematic uncertainty on the correction to go from the UPC to the photonuclear cross section. The mean value of \mant in each interval is also shown. 
}
\centering
\begin{tabular}{cccc}
\toprule
 Interval (GeV$^2 c^{-2}$) & $\left<|t|\right>$ (GeV$^2 c^{-2}$) &
$\frac{{\rm d}^{2}\sigma^{\text{coh}}_{\Jpsi}}{{\rm d}y{\rm d}\pttwo}$ ($\frac{{\rm mb} c^2}{{\rm GeV}^2}$)  &
$\frac{{\rm d}\sigma_{\gamma{\rm Pb}}}{{\rm d}|t|}$ ($\frac{{\rm mb} c^2}{{\rm GeV}^2}$) \\ 
\midrule
    $\left(0,0.72\right)\times10^{-3}$   & 0.00032 & 1290 $\pm 74 \pm 29 \pm 73$  & 8.15 $\pm 0.50 \pm 0.18 \pm 0.46 \pm 0.20$\\
    $\left(0.72,1.6\right)\times10^{-3}$ & 0.00113 & 1035 $\pm 47 \pm 10 \pm 60$  & 5.75 $\pm 0.27 \pm 0.06 \pm 0.34 \pm 0.16$\\
    $\left(1.6,2.6\right)\times10^{-3}$  & 0.00207 & 743  $\pm 34 \pm 6 \pm 43$   & 4.23 $\pm 0.20 \pm 0.03 \pm 0.25 \pm 0.11$\\
    $\left(2.6,4.0\right)\times10^{-3}$  & 0.00328 & 465  $\pm 24 \pm 6 \pm 27$   & 2.87 $\pm 0.15 \pm 0.04 \pm 0.17 \pm 0.08$\\
    $\left(4.0,6.2\right)\times10^{-3}$  & 0.00498 & 229  $\pm 14 \pm 3 \pm 14$   & 1.48 $\pm 0.09 \pm 0.02 \pm 0.09 \pm 0.04$\\
    $\left(6.2,12.1\right)\times10^{-3}$ & 0.00833 & 51   $\pm 5 \pm 1 \pm 4$     & 0.40 $\pm 0.04 \pm 0.01 \pm 0.03 \pm 0.03$\\
     \bottomrule
\end{tabular}
\label{tab_cs}
\end{table*}

\begin{figure}[tb]
    \begin{center}
    \includegraphics[width=0.95\linewidth]{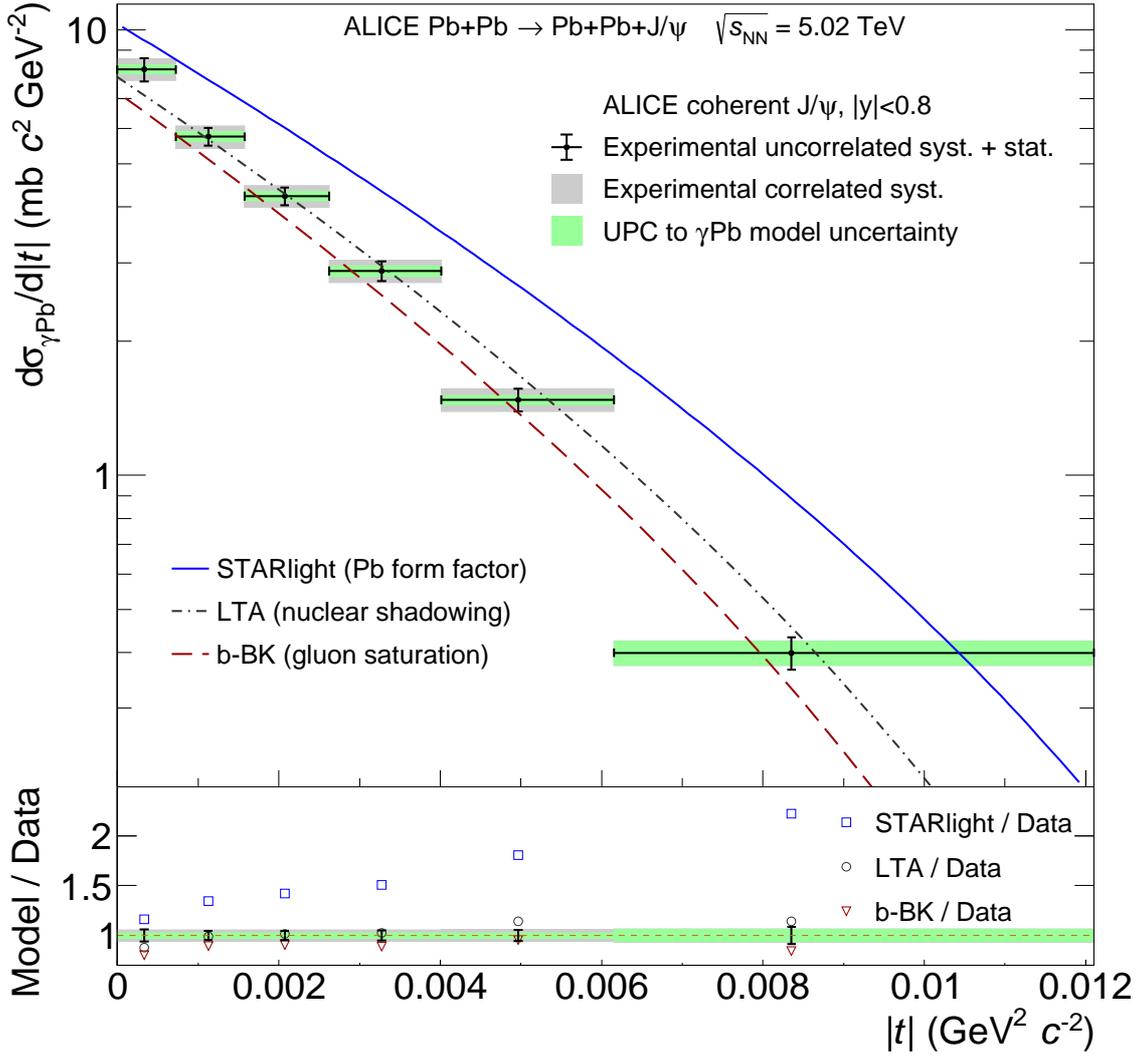}
    \end{center}
    \caption{Dependence on \mant of the photonuclear cross section for the coherent photoproduction of $\Jpsi$ off Pb compared with model predictions~\cite{Klein:2016yzr,Guzey:2016qwo,Bendova:2020hbb} (top panel), where for LTA  the {\em low shadowing} case is shown (see text). Model to data ratio for each prediction in each measured point (bottom panel). The uncertainties are split to those originating from experiment and to those originating from the correction to go from the UPC to the photonuclear cross section.}
    \label{fig_cs}
 \end{figure}

The results for the photonuclear cross section are listed in Table~\ref{tab_cs} and shown in Fig.~\ref{fig_cs}, where the measurement is compared with several theoretical predictions. The average \mant ($\left<|t|\right>$) quoted in Table~\ref{tab_cs} was estimated from the \mant-distribution used in the response matrix based on measured data (see above). The mean of the ensuing distribution in a given \pttwo interval was taken to be $\left<|t|\right>$.

STARlight utilises the vector meson dominance model and a parameterisation of the existing data on exclusive photoproduction of $\Jpsi$ off protons coupled with a Glauber-like formalism to obtain the photonuclear cross section. Since the \mant-dependence in this model comes from the Glauber calculation, meaning that it does not include explicitly gluon shadowing effects, it is an interesting baseline for comparisons (This approach is quite similar to the impulse approximation used in~\cite{Guzey:2013qza}). STARlight overestimates the measured cross section and the shape of the distribution appears to be wider than that of the measured data.

The LTA prediction by Guzey, Strikman and Zhalov~\cite{Guzey:2016qwo} is based on the leading-twist approximation (LTA) of nuclear shadowing based on the combination of the Gribov--Glauber theory and inclusive diffractive data from HERA~\cite{Frankfurt:2011cs}. There are  two LTA predictions; one called {\em high shadowing} and the other {\em low shadowing}. The low shadowing prediction is shown in Fig.~\ref{fig_cs}. The shape obtained from this model is similar to that of the data and describes the cross section within experimental uncertainties. As shown in Fig.~3 of~\cite{Guzey:2016qwo}, the high-shadowing version of the model has a similar shape but the overall normalisation is smaller by factor around 1.7.

The b-BK model by Bendova et al.~\cite{Cepila:2018faq,Bendova:2019psy,Bendova:2020hbb} is based on the colour dipole approach where the scattering amplitude is obtained from the impact-parameter dependent solution of the Balitsky--Kovchegov equation coupled to a nuclear-like intial condition~\cite{Balitsky:1995ub,Kovchegov:1999yj} which incorporates saturation effects. This model also predicts the behaviour of the data quite well.

The different predictions of the STARlight and LTA or b-BK models reflect the effects of QCD dynamics (shadowing in LTA, saturation in b-BK) at small values of $x\sim10^{-3}$ and highlight the importance of measuring the \mant-dependence of the photonuclear cross section.

\section{Conclusions}
\enlargethispage{\baselineskip}
The first measurement of the \mant-dependence of coherent \Jpsi photonuclear production off Pb nuclei in UPCs is presented. The measurement was carried out with the ALICE detector at midrapidity, $|y|<0.8$, in ultra-peripheral \PbPb collisions at \fivenn and covers the small-$x$ range  $(0.3-1.4)\times10^{-3}$. Photonuclear cross sections in six different intervals of \mant are reported and compared with theoretical predictions. The measured cross section shows a \mant-dependent shape different from a model based  on the Pb nuclear form factor and closer to the shape predicted by models including QCD dynamical effects in the form of shadowing (LTA) or saturation (b-BK). The difference in shape and magnitude between the LTA and b-BK models  is of the same order as the current measurement uncertainties, but the large data sample expected in the LHC Run 3~\cite{Citron:2018lsq} and the improvement in tracking from the upgrades of the ALICE detector~\cite{Abelevetal:2014cna} promise a much improved accuracy. These results highlight the importance of observables sensitive to the transverse gluonic structure of particles for extending the understanding of the high-energy limit of QCD.


\newenvironment{acknowledgement}{\relax}{\relax}
\begin{acknowledgement}
\section*{Acknowledgements}

The ALICE Collaboration would like to thank all its engineers and technicians for their invaluable contributions to the construction of the experiment and the CERN accelerator teams for the outstanding performance of the LHC complex.
The ALICE Collaboration gratefully acknowledges the resources and support provided by all Grid centres and the Worldwide LHC Computing Grid (WLCG) collaboration.
The ALICE Collaboration acknowledges the following funding agencies for their support in building and running the ALICE detector:
A. I. Alikhanyan National Science Laboratory (Yerevan Physics Institute) Foundation (ANSL), State Committee of Science and World Federation of Scientists (WFS), Armenia;
Austrian Academy of Sciences, Austrian Science Fund (FWF): [M 2467-N36] and Nationalstiftung f\"{u}r Forschung, Technologie und Entwicklung, Austria;
Ministry of Communications and High Technologies, National Nuclear Research Center, Azerbaijan;
Conselho Nacional de Desenvolvimento Cient\'{\i}fico e Tecnol\'{o}gico (CNPq), Financiadora de Estudos e Projetos (Finep), Funda\c{c}\~{a}o de Amparo \`{a} Pesquisa do Estado de S\~{a}o Paulo (FAPESP) and Universidade Federal do Rio Grande do Sul (UFRGS), Brazil;
Ministry of Education of China (MOEC) , Ministry of Science \& Technology of China (MSTC) and National Natural Science Foundation of China (NSFC), China;
Ministry of Science and Education and Croatian Science Foundation, Croatia;
Centro de Aplicaciones Tecnol\'{o}gicas y Desarrollo Nuclear (CEADEN), Cubaenerg\'{\i}a, Cuba;
Ministry of Education, Youth and Sports of the Czech Republic and Czech Science Foundation, Czech Republic;
The Danish Council for Independent Research | Natural Sciences, the VILLUM FONDEN and Danish National Research Foundation (DNRF), Denmark;
Helsinki Institute of Physics (HIP), Finland;
Commissariat \`{a} l'Energie Atomique (CEA) and Institut National de Physique Nucl\'{e}aire et de Physique des Particules (IN2P3) and Centre National de la Recherche Scientifique (CNRS), France;
Bundesministerium f\"{u}r Bildung und Forschung (BMBF) and GSI Helmholtzzentrum f\"{u}r Schwerionenforschung GmbH, Germany;
General Secretariat for Research and Technology, Ministry of Education, Research and Religions, Greece;
National Research, Development and Innovation Office, Hungary;
Department of Atomic Energy Government of India (DAE), Department of Science and Technology, Government of India (DST), University Grants Commission, Government of India (UGC) and Council of Scientific and Industrial Research (CSIR), India;
Indonesian Institute of Science, Indonesia;
Istituto Nazionale di Fisica Nucleare (INFN), Italy;
Institute for Innovative Science and Technology , Nagasaki Institute of Applied Science (IIST), Japanese Ministry of Education, Culture, Sports, Science and Technology (MEXT) and Japan Society for the Promotion of Science (JSPS) KAKENHI, Japan;
Consejo Nacional de Ciencia (CONACYT) y Tecnolog\'{i}a, through Fondo de Cooperaci\'{o}n Internacional en Ciencia y Tecnolog\'{i}a (FONCICYT) and Direcci\'{o}n General de Asuntos del Personal Academico (DGAPA), Mexico;
Nederlandse Organisatie voor Wetenschappelijk Onderzoek (NWO), Netherlands;
The Research Council of Norway, Norway;
Commission on Science and Technology for Sustainable Development in the South (COMSATS), Pakistan;
Pontificia Universidad Cat\'{o}lica del Per\'{u}, Peru;
Ministry of Science and Higher Education, National Science Centre and WUT ID-UB, Poland;
Korea Institute of Science and Technology Information and National Research Foundation of Korea (NRF), Republic of Korea;
Ministry of Education and Scientific Research, Institute of Atomic Physics and Ministry of Research and Innovation and Institute of Atomic Physics, Romania;
Joint Institute for Nuclear Research (JINR), Ministry of Education and Science of the Russian Federation, National Research Centre Kurchatov Institute, Russian Science Foundation and Russian Foundation for Basic Research, Russia;
Ministry of Education, Science, Research and Sport of the Slovak Republic, Slovakia;
National Research Foundation of South Africa, South Africa;
Swedish Research Council (VR) and Knut \& Alice Wallenberg Foundation (KAW), Sweden;
European Organization for Nuclear Research, Switzerland;
Suranaree University of Technology (SUT), National Science and Technology Development Agency (NSDTA) and Office of the Higher Education Commission under NRU project of Thailand, Thailand;
Turkish Atomic Energy Agency (TAEK), Turkey;
National Academy of  Sciences of Ukraine, Ukraine;
Science and Technology Facilities Council (STFC), United Kingdom;
National Science Foundation of the United States of America (NSF) and United States Department of Energy, Office of Nuclear Physics (DOE NP), United States of America.
\end{acknowledgement}

\bibliographystyle{utphys}   
\bibliography{bibliography}

\newpage
\appendix

%
%

\section{The ALICE Collaboration}
\label{app:collab}
\begingroup
\small
\begin{flushleft} 


S.~Acharya$^{\rm 142}$, 
D.~Adamov\'{a}$^{\rm 97}$, 
A.~Adler$^{\rm 75}$, 
J.~Adolfsson$^{\rm 82}$, 
G.~Aglieri Rinella$^{\rm 35}$, 
M.~Agnello$^{\rm 31}$, 
N.~Agrawal$^{\rm 55}$, 
Z.~Ahammed$^{\rm 142}$, 
S.~Ahmad$^{\rm 16}$, 
S.U.~Ahn$^{\rm 77}$
% ...

\section*{Affiliation notes}

$^{\rm I}$ Deceased\\
$^{\rm II}$ Also at: Italian National Agency for New Technologies, Energy and Sustainable Economic Development (ENEA), Bologna, Italy\\
$^{\rm III}$ Also at: Dipartimento DET del Politecnico di Torino, Turin, Italy\\
$^{\rm IV}$ Also at: M.V. Lomonosov Moscow State University, D.V. Skobeltsyn Institute of Nuclear, Physics, Moscow, Russia\\
$^{\rm V}$ Also at: Institute of Theoretical Physics, University of Wroclaw, Poland\\

\section*{Collaboration Institutes}

$^{1}$ A.I. Alikhanyan National Science Laboratory (Yerevan Physics Institute) Foundation, Yerevan, Armenia\\
$^{2}$ AGH University of Science and Technology, Cracow, Poland\\
$^{3}$ Bogolyubov Institute for Theoretical Physics, National Academy of Sciences of Ukraine, Kiev, Ukraine\\
$^{4}$ Bose Institute, Department of Physics  and Centre for Astroparticle Physics and Space Science (CAPSS), Kolkata, India\\
$^{5}$ Budker Institute for Nuclear Physics, Novosibirsk, Russia\\
$^{6}$ California Polytechnic State University, San Luis Obispo, California, United States\\
$^{7}$ Central China Normal University, Wuhan, China\\
$^{8}$ Centro de Aplicaciones Tecnol\'{o}gicas y Desarrollo Nuclear (CEADEN), Havana, Cuba\\
$^{9}$ Centro de Investigaci\'{o}n y de Estudios Avanzados (CINVESTAV), Mexico City and M\'{e}rida, Mexico\\
$^{10}$ Chicago State University, Chicago, Illinois, United States\\
$^{11}$ China Institute of Atomic Energy, Beijing, China\\
$^{12}$ Chungbuk National University, Cheongju, Republic of Korea\\
$^{13}$ Comenius University Bratislava, Faculty of Mathematics, Physics and Informatics, Bratislava, Slovakia\\
$^{14}$ COMSATS University Islamabad, Islamabad, Pakistan\\
$^{15}$ Creighton University, Omaha, Nebraska, United States\\
$^{16}$ Department of Physics, Aligarh Muslim University, Aligarh, India\\
$^{17}$ Department of Physics, Pusan National University, Pusan, Republic of Korea\\
$^{18}$ Department of Physics, Sejong University, Seoul, Republic of Korea\\
$^{19}$ Department of Physics, University of California, Berkeley, California, United States\\
$^{20}$ Department of Physics, University of Oslo, Oslo, Norway\\
$^{21}$ Department of Physics and Technology, University of Bergen, Bergen, Norway\\
$^{22}$ Dipartimento di Fisica dell'Universit\`{a} 'La Sapienza' and Sezione INFN, Rome, Italy\\
$^{23}$ Dipartimento di Fisica dell'Universit\`{a} and Sezione INFN, Cagliari, Italy\\
$^{24}$ Dipartimento di Fisica dell'Universit\`{a} and Sezione INFN, Trieste, Italy\\
$^{25}$ Dipartimento di Fisica dell'Universit\`{a} and Sezione INFN, Turin, Italy\\
$^{26}$ Dipartimento di Fisica e Astronomia dell'Universit\`{a} and Sezione INFN, Bologna, Italy\\
$^{27}$ Dipartimento di Fisica e Astronomia dell'Universit\`{a} and Sezione INFN, Catania, Italy\\
$^{28}$ Dipartimento di Fisica e Astronomia dell'Universit\`{a} and Sezione INFN, Padova, Italy\\
$^{29}$ Dipartimento di Fisica e Nucleare e Teorica, Universit\`{a} di Pavia  and Sezione INFN, Pavia, Italy\\
$^{30}$ Dipartimento di Fisica `E.R.~Caianiello' dell'Universit\`{a} and Gruppo Collegato INFN, Salerno, Italy\\
$^{31}$ Dipartimento DISAT del Politecnico and Sezione INFN, Turin, Italy\\
$^{32}$ Dipartimento di Scienze e Innovazione Tecnologica dell'Universit\`{a} del Piemonte Orientale and INFN Sezione di Torino, Alessandria, Italy\\
$^{33}$ Dipartimento di Scienze MIFT, Universit\`{a} di Messina, Messina, Italy\\
$^{34}$ Dipartimento Interateneo di Fisica `M.~Merlin' and Sezione INFN, Bari, Italy\\
$^{35}$ European Organization for Nuclear Research (CERN), Geneva, Switzerland\\
$^{36}$ Faculty of Electrical Engineering, Mechanical Engineering and Naval Architecture, University of Split, Split, Croatia\\
$^{37}$ Faculty of Engineering and Science, Western Norway University of Applied Sciences, Bergen, Norway\\
$^{38}$ Faculty of Nuclear Sciences and Physical Engineering, Czech Technical University in Prague, Prague, Czech Republic\\
$^{39}$ Faculty of Science, P.J.~\v{S}af\'{a}rik University, Ko\v{s}ice, Slovakia\\
$^{40}$ Frankfurt Institute for Advanced Studies, Johann Wolfgang Goethe-Universit\"{a}t Frankfurt, Frankfurt, Germany\\
$^{41}$ Fudan University, Shanghai, China\\
$^{42}$ Gangneung-Wonju National University, Gangneung, Republic of Korea\\
$^{43}$ Gauhati University, Department of Physics, Guwahati, India\\
$^{44}$ Helmholtz-Institut f\"{u}r Strahlen- und Kernphysik, Rheinische Friedrich-Wilhelms-Universit\"{a}t Bonn, Bonn, Germany\\
$^{45}$ Helsinki Institute of Physics (HIP), Helsinki, Finland\\
$^{46}$ High Energy Physics Group,  Universidad Aut\'{o}noma de Puebla, Puebla, Mexico\\
$^{47}$ Hiroshima University, Hiroshima, Japan\\
$^{48}$ Hochschule Worms, Zentrum  f\"{u}r Technologietransfer und Telekommunikation (ZTT), Worms, Germany\\
$^{49}$ Horia Hulubei National Institute of Physics and Nuclear Engineering, Bucharest, Romania\\
$^{50}$ Indian Institute of Technology Bombay (IIT), Mumbai, India\\
$^{51}$ Indian Institute of Technology Indore, Indore, India\\
$^{52}$ Indonesian Institute of Sciences, Jakarta, Indonesia\\
$^{53}$ INFN, Laboratori Nazionali di Frascati, Frascati, Italy\\
$^{54}$ INFN, Sezione di Bari, Bari, Italy\\
$^{55}$ INFN, Sezione di Bologna, Bologna, Italy\\
$^{56}$ INFN, Sezione di Cagliari, Cagliari, Italy\\
$^{57}$ INFN, Sezione di Catania, Catania, Italy\\
$^{58}$ INFN, Sezione di Padova, Padova, Italy\\
$^{59}$ INFN, Sezione di Roma, Rome, Italy\\
$^{60}$ INFN, Sezione di Torino, Turin, Italy\\
$^{61}$ INFN, Sezione di Trieste, Trieste, Italy\\
$^{62}$ Inha University, Incheon, Republic of Korea\\
$^{63}$ Institute for Gravitational and Subatomic Physics (GRASP), Utrecht University/Nikhef, Utrecht, Netherlands\\
$^{64}$ Institute for Nuclear Research, Academy of Sciences, Moscow, Russia\\
$^{65}$ Institute of Experimental Physics, Slovak Academy of Sciences, Ko\v{s}ice, Slovakia\\
$^{66}$ Institute of Physics, Homi Bhabha National Institute, Bhubaneswar, India\\
$^{67}$ Institute of Physics of the Czech Academy of Sciences, Prague, Czech Republic\\
$^{68}$ Institute of Space Science (ISS), Bucharest, Romania\\
$^{69}$ Institut f\"{u}r Kernphysik, Johann Wolfgang Goethe-Universit\"{a}t Frankfurt, Frankfurt, Germany\\
$^{70}$ Instituto de Ciencias Nucleares, Universidad Nacional Aut\'{o}noma de M\'{e}xico, Mexico City, Mexico\\
$^{71}$ Instituto de F\'{i}sica, Universidade Federal do Rio Grande do Sul (UFRGS), Porto Alegre, Brazil\\
$^{72}$ Instituto de F\'{\i}sica, Universidad Nacional Aut\'{o}noma de M\'{e}xico, Mexico City, Mexico\\
$^{73}$ iThemba LABS, National Research Foundation, Somerset West, South Africa\\
$^{74}$ Jeonbuk National University, Jeonju, Republic of Korea\\
$^{75}$ Johann-Wolfgang-Goethe Universit\"{a}t Frankfurt Institut f\"{u}r Informatik, Fachbereich Informatik und Mathematik, Frankfurt, Germany\\
$^{76}$ Joint Institute for Nuclear Research (JINR), Dubna, Russia\\
$^{77}$ Korea Institute of Science and Technology Information, Daejeon, Republic of Korea\\
$^{78}$ KTO Karatay University, Konya, Turkey\\
$^{79}$ Laboratoire de Physique des 2 Infinis, Ir\`{e}ne Joliot-Curie, Orsay, France\\
$^{80}$ Laboratoire de Physique Subatomique et de Cosmologie, Universit\'{e} Grenoble-Alpes, CNRS-IN2P3, Grenoble, France\\
$^{81}$ Lawrence Berkeley National Laboratory, Berkeley, California, United States\\
$^{82}$ Lund University Department of Physics, Division of Particle Physics, Lund, Sweden\\
$^{83}$ Moscow Institute for Physics and Technology, Moscow, Russia\\
$^{84}$ Nagasaki Institute of Applied Science, Nagasaki, Japan\\
$^{85}$ Nara Women{'}s University (NWU), Nara, Japan\\
$^{86}$ National and Kapodistrian University of Athens, School of Science, Department of Physics , Athens, Greece\\
$^{87}$ National Centre for Nuclear Research, Warsaw, Poland\\
$^{88}$ National Institute of Science Education and Research, Homi Bhabha National Institute, Jatni, India\\
$^{89}$ National Nuclear Research Center, Baku, Azerbaijan\\
$^{90}$ National Research Centre Kurchatov Institute, Moscow, Russia\\
$^{91}$ Niels Bohr Institute, University of Copenhagen, Copenhagen, Denmark\\
$^{92}$ Nikhef, National institute for subatomic physics, Amsterdam, Netherlands\\
$^{93}$ NRC Kurchatov Institute IHEP, Protvino, Russia\\
$^{94}$ NRC \guillemotleft Kurchatov\guillemotright  Institute - ITEP, Moscow, Russia\\
$^{95}$ NRNU Moscow Engineering Physics Institute, Moscow, Russia\\
$^{96}$ Nuclear Physics Group, STFC Daresbury Laboratory, Daresbury, United Kingdom\\
$^{97}$ Nuclear Physics Institute of the Czech Academy of Sciences, \v{R}e\v{z} u Prahy, Czech Republic\\
$^{98}$ Oak Ridge National Laboratory, Oak Ridge, Tennessee, United States\\
$^{99}$ Ohio State University, Columbus, Ohio, United States\\
$^{100}$ Petersburg Nuclear Physics Institute, Gatchina, Russia\\
$^{101}$ Physics department, Faculty of science, University of Zagreb, Zagreb, Croatia\\
$^{102}$ Physics Department, Panjab University, Chandigarh, India\\
$^{103}$ Physics Department, University of Jammu, Jammu, India\\
$^{104}$ Physics Department, University of Rajasthan, Jaipur, India\\
$^{105}$ Physikalisches Institut, Eberhard-Karls-Universit\"{a}t T\"{u}bingen, T\"{u}bingen, Germany\\
$^{106}$ Physikalisches Institut, Ruprecht-Karls-Universit\"{a}t Heidelberg, Heidelberg, Germany\\
$^{107}$ Physik Department, Technische Universit\"{a}t M\"{u}nchen, Munich, Germany\\
$^{108}$ Politecnico di Bari and Sezione INFN, Bari, Italy\\
$^{109}$ Research Division and ExtreMe Matter Institute EMMI, GSI Helmholtzzentrum f\"ur Schwerionenforschung GmbH, Darmstadt, Germany\\
$^{110}$ Rudjer Bo\v{s}kovi\'{c} Institute, Zagreb, Croatia\\
$^{111}$ Russian Federal Nuclear Center (VNIIEF), Sarov, Russia\\
$^{112}$ Saha Institute of Nuclear Physics, Homi Bhabha National Institute, Kolkata, India\\
$^{113}$ School of Physics and Astronomy, University of Birmingham, Birmingham, United Kingdom\\
$^{114}$ Secci\'{o}n F\'{\i}sica, Departamento de Ciencias, Pontificia Universidad Cat\'{o}lica del Per\'{u}, Lima, Peru\\
$^{115}$ St. Petersburg State University, St. Petersburg, Russia\\
$^{116}$ Stefan Meyer Institut f\"{u}r Subatomare Physik (SMI), Vienna, Austria\\
$^{117}$ SUBATECH, IMT Atlantique, Universit\'{e} de Nantes, CNRS-IN2P3, Nantes, France\\
$^{118}$ Suranaree University of Technology, Nakhon Ratchasima, Thailand\\
$^{119}$ Technical University of Ko\v{s}ice, Ko\v{s}ice, Slovakia\\
$^{120}$ The Henryk Niewodniczanski Institute of Nuclear Physics, Polish Academy of Sciences, Cracow, Poland\\
$^{121}$ The University of Texas at Austin, Austin, Texas, United States\\
$^{122}$ Universidad Aut\'{o}noma de Sinaloa, Culiac\'{a}n, Mexico\\
$^{123}$ Universidade de S\~{a}o Paulo (USP), S\~{a}o Paulo, Brazil\\
$^{124}$ Universidade Estadual de Campinas (UNICAMP), Campinas, Brazil\\
$^{125}$ Universidade Federal do ABC, Santo Andre, Brazil\\
$^{126}$ University of Cape Town, Cape Town, South Africa\\
$^{127}$ University of Houston, Houston, Texas, United States\\
$^{128}$ University of Jyv\"{a}skyl\"{a}, Jyv\"{a}skyl\"{a}, Finland\\
$^{129}$ University of Liverpool, Liverpool, United Kingdom\\
$^{130}$ University of Science and Technology of China, Hefei, China\\
$^{131}$ University of South-Eastern Norway, Tonsberg, Norway\\
$^{132}$ University of Tennessee, Knoxville, Tennessee, United States\\
$^{133}$ University of the Witwatersrand, Johannesburg, South Africa\\
$^{134}$ University of Tokyo, Tokyo, Japan\\
$^{135}$ University of Tsukuba, Tsukuba, Japan\\
$^{136}$ Universit\'{e} Clermont Auvergne, CNRS/IN2P3, LPC, Clermont-Ferrand, France\\
$^{137}$ Universit\'{e} de Lyon, CNRS/IN2P3, Institut de Physique des 2 Infinis de Lyon , Lyon, France\\
$^{138}$ Universit\'{e} de Strasbourg, CNRS, IPHC UMR 7178, F-67000 Strasbourg, France, Strasbourg, France\\
$^{139}$ Universit\'{e} Paris-Saclay Centre d'Etudes de Saclay (CEA), IRFU, D\'{e}partment de Physique Nucl\'{e}aire (DPhN), Saclay, France\\
$^{140}$ Universit\`{a} degli Studi di Foggia, Foggia, Italy\\
$^{141}$ Universit\`{a} di Brescia and Sezione INFN, Brescia, Italy\\
$^{142}$ Variable Energy Cyclotron Centre, Homi Bhabha National Institute, Kolkata, India\\
$^{143}$ Warsaw University of Technology, Warsaw, Poland\\
$^{144}$ Wayne State University, Detroit, Michigan, United States\\
$^{145}$ Westf\"{a}lische Wilhelms-Universit\"{a}t M\"{u}nster, Institut f\"{u}r Kernphysik, M\"{u}nster, Germany\\
$^{146}$ Wigner Research Centre for Physics, Budapest, Hungary\\
$^{147}$ Yale University, New Haven, Connecticut, United States\\
$^{148}$ Yonsei University, Seoul, Republic of Korea\\

\bigskip 

\end{flushleft} 
\endgroup
  
\end{document}